# ESTABLISHMENT AND PLACEMENT OF A MULTI-PURPOSE PHASOR MEASUREMENT UNIT TO IMPROVE PARALLEL STATE ESTIMATION IN DISTRIBUTION NETWORKS

Zeyad Khashroum[1], Ali Dehghan Chaharabi[2], Lorena Palmero[3], Keiichiro Yasukawa[4]

[1] Department of Electrical Engineering, Lamar University, USA
[2] Department of Electrical Engineering, Bushehr Branch, Islamic Azad University, Iran
[3] Department of Electrical Engineering, University of Seville, Spain
[4] Department of Electrical Engineering, University of Pavia, Italy

*Corresponding Author: Zeyad Khashroum
Corresponding Author Email: zeiad89@gmail.com
**Article Received:** 10-09-21   **Accepted:** 22-09-21   **Published:** 30-09-21

## ABSTRACT

Today, microgrids in distribution networks are in dire need of improvement to cope with economic challenges, human losses, and equipment placement issues. Today, there is the issue of scattered resources in distribution systems, which has created many problems in the areas of environment, economy, and human and animal losses. The most important challenge in this section is the existence of voltage and frequency fluctuations during the occurrence of possible events such as severe load changes or errors in distribution networks. Having such a big problem can call a distribution network into question and destroy it. Therefore, it is necessary to provide an optimal method that can meet and cover these challenges. For this purpose, the present research deals with the problem of establishing and placing a multifunctional phasor measurement unit to improve the parallel state estimation in distribution networks, which offers a control approach. This approach determines the time of occurrence of internal and external disturbances after using the phasor unit. The approach of this research is to use a neural-fuzzy method because there is uncertainty in the distribution network due to the mentioned challenges, and training in the system is needed to accurately deploy and place possible errors. Do not occur. When setting up and placing the phasor measuring unit, the most important issue is the proper distribution of the load in the distribution network. The simulation results in the MATLAB / Simulink environment show the improvement of the results according to the proposed approach.

## INTRODUCTION

Global warming and environmental problems such as landfills and other sources of air, water, and land pollutions are highly considered around the World. Researchers are trying to pave the way for managing and removing these pollutions and using clean energies as much as possible (Nochian, Mohd Tahir, Mualan, & Rui, 2019). Power systems suffer from many problems such as increasing load demand, losses, limitation of generation, power supply issues, safety, voltage stability, management of density (Hayerikhiyavi & Dimitrovski, 2021a; Jasni & Ab Kadir, 2011). Furthermore, a considerable portion of electric energy consumes by electrical motors in industrial, domestic, and public usages. Therefore, improve the efficiency of electrical motors is another main issue in the power industry (Karami, Mariun, Ab-Kadir, Misron, & Mohd Radzi, 2021). The aforementioned problems attract attention around the World and many research teams try to propose effective methods as a solution. Microgrids are one of the interesting solutions for power system problems.

Microgrids are becoming an effective way to solve the power supply problem on off-grid islands. Investment economics is one of the main factors influencing its development and application, which is very challenging due to various vague information involved in the investment decision-making process. The presence of scattered resources and micro-grids in the power system, despite the many economic and environmental benefits, has added new problems to the power system. These problems include voltage and frequency fluctuations when possible events occur, such as severe load changes or errors in the power system. In the case of high impact low probability events, due to the lack of backup power, the intensity and amplitude of these fluctuations and the possibility of instability and collapse of the microgrid are much higher (Haggi, Song, & Sun, 2019). In addition, due to the low inertia of the scattered resources in the microgrid and the high switching speed of power electronics, the dynamics of an island microgrid are much faster than conventional power systems. Therefore, it is necessary to have an efficient control structure with fast performance when a disturbance occurs in the system.

Traditional distribution networks frequency and voltage deviations are considered as an indicator for detection of security systems through phasor measurement units (Haggi, Sun, & Qi, 2020; Hayerikhiyavi & Dimitrovski, 2021). Whereas in an independent micro-grid with resources with an electronic power interface, frequency and voltage deviations in the micro-grid due to power and load disturbances are well controlled by the real power frequency controller and the voltage-reactive power controller. What happens in traditional distribution networks, frequency and voltage deviations are not an important issue for the security detection of independent micro-grids with resources with electronic power interface. But in an independent micro-grid, the balance between power generation and power consumption is considered as an indicator to assess its safety, especially in the event of power and load disturbances. Therefore, if this balance is not created in the micro-grid, the microgrid is considered unsafe. In this case, the need for preventive measures such as load shedding and regulation of production in the shortest time in the microgrid should be done (Agarwal, Niknejad, Rahmani, Barzegaran, & Vanfretti, 2021). In this research, a new strategy for investigating the dynamic security of micro-grids is proposed to investigate this issue as a short-term planning decision model and the occurrence of power and load disorders as well as the use of adaptive fuzzy neural networks are examined evolutionarily. Also, the case of detection of micro-grid insecurity, the prediction of preventive control actions by other neural networks is suggested (Lasseter, 2011; Lasseter & Paigi, 2004).

## LITERATURE REVIEW

Microgrid is a collection of scattered and load-bearing production resources. Various categories have been proposed for energy sources in microgrids. One of these is a classification based on how energy sources are connected to the microgrid. Accordingly, one group of units is connected to the microgrid via a synchronous machine, and the other group is those that provide the electronic power interface to connect to the microgrid. In microgrids with distributed generation resources with electronic power interface for operation in transient network mode or changes especially in (DS) load demand, usually, an energy storage system is required when the microgrid is operated in stand-alone island mode (Aktarujjaman, Haque, Muttaqi, Negnevitsky, & Ledwich, 2007; Farhangi, 2009; Katiraei, Iravani, Hatziargyriou, & Dimeas, 2008; Lasseter, 2002; Miao, Domijan, & Fan, 2011; Peng, Li, & Tolbert, 2009). The microgrid can be used in two modes connected to the network and the independent island in the mode of connection to the main network. But when due to voltage drops, errors, blackouts, etc. The microgrid is slowly transferred to the operation of an independent island, the power balance in a separate microgrid, especially in the event of power and load disturbances, is a vital issue to continue the safe operation of the microgrid. Arises. If in the case of an independent island, due to power disturbances or increased load consumption, the distributed generation resources in the microgrid are not able to provide power, the energy storage system can be used, but due to its limitation, if necessary, it should be reduced by 5 times. But the least amount of load loss and reaching the power balance state in the microgrid faster is an issue that should be achieved (Chandorkar, Divan, & Adapa, 1993; Ghazanfari, Hamzeh, Mokhtari, & Karimi, 2012; Piagi & Lasseter, 2006). In the microgrid to achieve the desired installation and withdrawal characteristics of resources, frequency and voltage drop control are used to adjust the actual and reactive power. In this method, the share of each source is obtained through the inverter interface based on the drop curve characteristic, which causes a quick response and by assigning the reference amount to each unit, prevents damage to the scattered products. Frequency deviation can be limited by defining the frequency drop characteristic and even returning it to the nominal value using the frequency reduction loop. Also, using the voltage drop characteristic, terminal voltage changes are limited. As a result, distributed generation units with power electronic interfaces react to voltage deviations due to changes in microgrid or local load within the allowable range. So with this method, the frequency and voltage of the microgrid can be adjusted so that they are prevented from leaving the allowable value. This means that, unlike traditional systems in microgrids, frequency and voltage deviations are not considered as an indicator to assess microgrid security. Thus, it is important to find methods for rapid detection of independent microgrid security, especially in the event of power and load disturbances. Also, if the microgrid is unsafe, immediate action is necessary for the safe operation of the microgrid, which should be investigated. To evaluate microgrid security traditionally, the most accurate method is to solve a set of nonlinear equations, which is a very difficult and time-consuming computational method. But the use of artificial intelligence tools is a good alternative to quickly and accurately describe microgrid security (Chandorkar et al., 1993).

The frequency deviation and storage equipment performance in microgrids is examined (Ghazanfari et al., 2012; Hamzeh, Mokhtari, & Karimi, 2013; Katiraei & Iravani, 2006; Lopes, Moreira, & Madureira, 2006) and indicators to assess the security of the transfer of the microgrid distinct modes when unplanned network disruption medium voltage high is provided. This article (Moreira & Lopes, 2007) emphasizes the use of artificial neural networks due to their computational speed in online performance and their flexibility to predict corrective actions in unsafe operating modes to achieve a smooth transition between connected and separate performance. To evaluate the security of the traditional distribution network, voltage deviation has been investigated. In (Jasni & Ab Kadir, 2011), an artificial neural network has been used to evaluate the security of the standard 9-bus network. When a system is designed with only artificial neural networks, the network is a black box that needs to be defined. This is a very computational and heavy process. After extensive experience and practice on the complexity of the network and the learning algorithm to use and the degree of accuracy that is acceptable in this application, the designer can achieve relative satisfaction. If we incorporate the functions of fuzzy logic into neural networks and learning and share neural network classification in fuzzy systems, then the shortcomings of neural networks and fuzzy systems can be covered. The result will be an adaptive neural-fuzzy network.

In adaptive neural-fuzzy networks, first, the neural network part is used to learn it and classify the abilities and to link the pattern and modify the pattern. The neural network part automatically creates the rules of fuzzy logic and membership functions during the learning cycle. In general, even after learning, the neural network continues to modify membership functions and fuzzy logic rules, learning more and more from its input signals. Fuzzy logic, on the other hand, is used to infer and provide a definite or non-fuzzy output (when fuzzy variables are generated). In general, microgrids are systems that arise from the integration of distributed generation units, energy storage systems, and controllable loads in low-voltage and medium-voltage networks and can be operated in either grid-connected state or independently to be. Microgrids have many benefits, including improved power quality and reliability, reduced losses, economic benefits, and reduced environmental pollution. In recent years, electric vehicles as an energy storage system as well as a public vehicle have undergone significant development and progress, which has been considered due to the need for fossil fuels and the reduction of environmental pollution. According to the forecasts, the existence of electric vehicles as an emerging phenomenon in the electricity network should be investigated. Each microgrid can exchange power with other microgrids and the main network or buy and sell power in some way. This will increase reliability as well as help maintain a balance between supply and demand. Energy sources in microgrids consist of units such as wind power plants, photovoltaic power plants that have uncertainty in power generation and units such as microturbines, fuel cells, heat and power plants that have pollution (Jang, 1993). In (Liu, Guo, Hou, Wang, & Wang, 2021) a randomized multi-period investment planning model is presented for the islanded microgrid. The application of the model in dealing with various uncertainties is enhanced through a hybrid optimization framework in which long-term uncertainty of energy price fluctuations is captured by a random programming approach and short-term changes in production and renewable energy load are considered. Dynamic information from load growth, unit cost change and device demolition is considered to make the decision more practical and economically attractive. The multi-period investment planning model is formulated as a complex integer linear programming problem, and decision conservatism can be flexibly adjusted by adjusting the power of the model. The simulation results based on real data show that the proposed model shows better economic performance and synergy than the traditional multi-year optimization model, with the total planning cost decreased by almost 3.6% and the initial investment cost decreased by approximately 36.6% and the use of renewable energy increased by approximately 5.4%. In addition, sensitivity analysis for load growth rate, loan ratio, unit cost and uncertain budgets further confirms the application of the proposed model under different conditions (Rahmani, Robinson, & Barzegaran, 2021).

## PROPOSED MODEL

In this research, a new strategy to investigate the dynamic security of the distribution network in a part of the microgrid is proposed as a model for establishing a phasor measurement unit based on parallel state estimation and the occurrence of power and load disorders as well as the use of adaptive neural-fuzzy networks. Evolutionary is examined. Also, in case of detection of microgrid insecurity, it is recommended to anticipate preventive control actions by other neural networks. First, in Table 1, the names that are used as abbreviations in pictures and relationships are stated. Figure 1 shows the initial configuration of the microgrid in the distribution network.

**Table 1: Abbreviated Nouns**

| | |
|---|---|
| **APF** | active power filter |
| **ANFIS** | adaptive neuro fuzzy inference |
| **MG** | microgrid system |
| **MGU** | microgrid system utility |
| **PQ** | power quality |
| **PCC** | point of common coupling |
| **UPFC** | unified power quality conditioner |
| **VSI** | voltage source inverter |

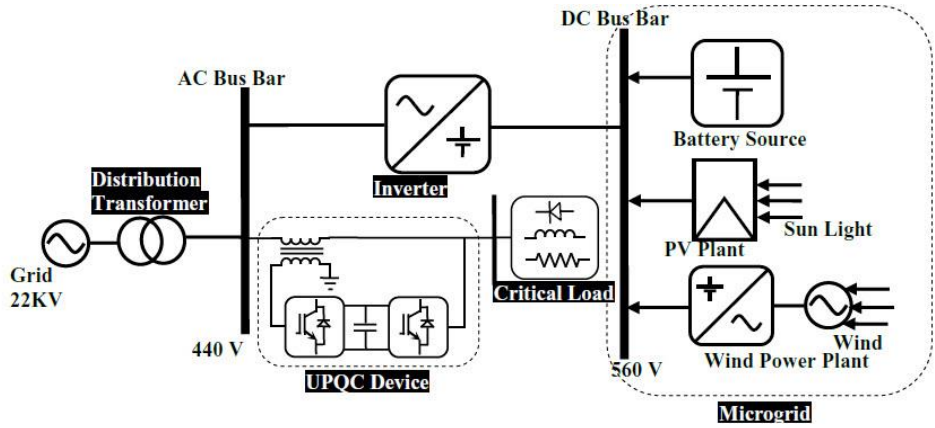

Figure 1: Initial Microgrid Configuration in the Distribution Network

A hybrid microgrid in a distribution network is made up of several alternating energy sources connected to a common bus. The microgrid structure must be clear. All power sources are connected to the DC bus. The energy generated is connected to DC, which is done using a rectifier. In assistive devices, the AC bus is connected to a distribution converter. The DC bus is connected to the AC bus using a power inverter. The critical load is also connected to the AC bus using the UPFC device. The UPFC alignment power circuit shown in Figure 2 is shown.

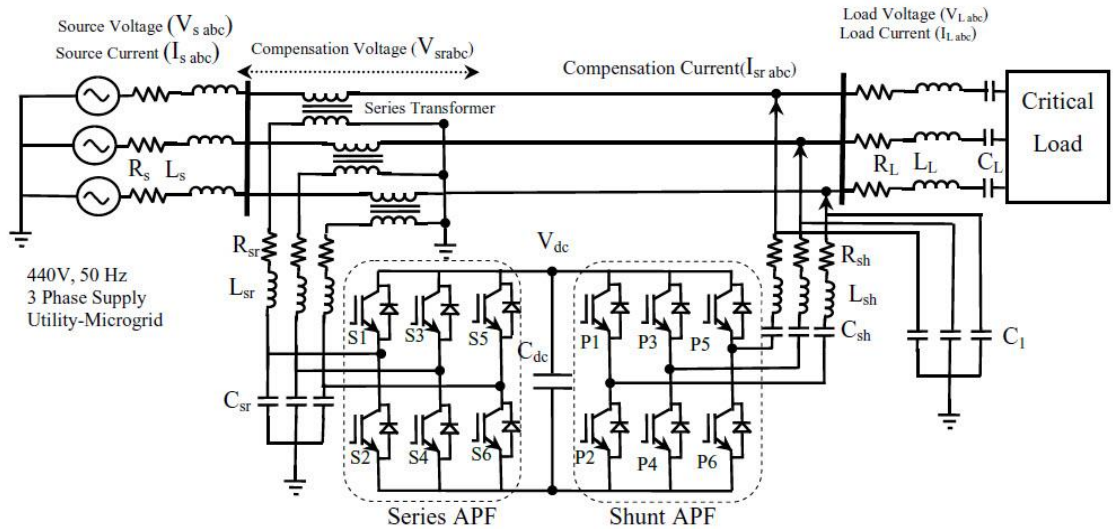

Figure 2: UPFC Construction

The UPFC alignment is shared or designed from two voltage source inverters, or VSIs (Voltage Source Inverters), with a DC link capacitor. A VSI is connected in series with the transformer connection source, whose functions are in the form of an APF series. APF series is responsible for compensating for PQ distortion at common (Point of Common Coupling) points. Other VSIs are connected in parallel, located between the APF series and the load acting as the APF shunt. The shunt is responsible for compensating for power quality problems associated with customers and regulating the DC connection voltage. In the alignment presented in the short-term decision model, shunt passive series capacitors or C_sh play an important role in supporting shunt APF. In the loading section, the passive series capacitor or C_L supports reactive power flow in pairing. The main purpose of the control technique presented in this research is to adjust both series and APF shunts for optimal power distribution in the distribution network microgrids for the establishment and placement of the phasor measurement unit based on parallel state estimation. Therefore, the microgrid control technique in this research is classified into series and shunt control techniques, which will eventually be combined with neural-fuzzy networks. The series control technique is designed to adjust the load rates in model T to establish and place the phasor measurement unit based on parallel state estimation. The block diagram of the series control technique is shown in Figure 3.

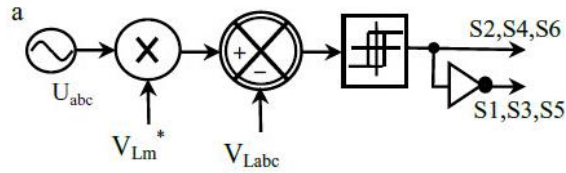

Figure 3: Series Control Technique

The series control technique regulates the load voltage using a process such as generating a reference signal and works by recording PQ tuning instruments and generating pulses for VSI series. Reference load voltage from the output voltage size or $V_{Lm}$ and also the sinusoidal signal of the phasor unit or is formed, the relation of which is in the form of equation (1).

$$\begin{bmatrix} V_{La}^* \\ V_{Lb}^* \\ V_{Lc}^* \end{bmatrix} = \begin{bmatrix} V_{Lm}^* \sin(\omega t) \\ V_{Lm}^* \sin(\omega t + 120) \\ V_{Lm}^* = \sin(\omega t - 120) \end{bmatrix} \quad (1)$$

According to Equation (1), the number 120 means 120 °C. The PQ distortion instrument is recorded from various sources and the actual load voltage with the help of Equation (2). Pulses for VSI series are generated from the perturbation instrument using a hysteresis controller.

$$\begin{bmatrix} V_{Ca} \\ V_{Cb} \\ V_{cb} \end{bmatrix} = \begin{bmatrix} V_{La}^* \\ V_{Lb}^* \\ V_{Lc}^* \end{bmatrix} - \begin{bmatrix} V_{La} \\ V_{Lb} \\ V_{Lc} \end{bmatrix} \quad (2)$$

Proposed shunt control technique to adjust the current source and connection voltage or $V_{dc}$ designed. The block diagram of the shunt control technique is shown in Figure 4.

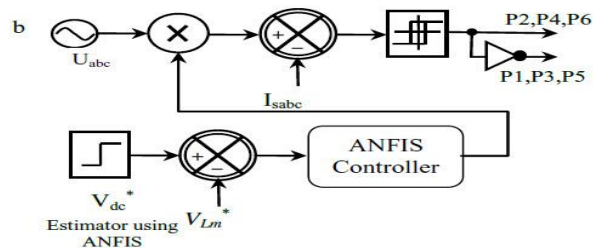

Figure 4: Shunt Control Technique

The purpose function of APF shunts, such as the process of estimating the reference DC connection voltage or $V_{dc}^*$, controller $V_{dc}$, Reference source output generation, PQ distortion recording is obtained from customers and pulses are generated for VSI shunt. In the proposed technique, neural-fuzzy network or ANFIS, for estimation $V_{dc}^*$, controller $V_{dc}$, Trained. The ANFIS controller is trained to generate a base of real resource compensation components. The reference source current generates the base production of the actual components of the source current and the three-phasor sine vector with the help of Equation (3).

$$\begin{bmatrix} I_{sa}^* \\ I_{sb}^* \\ I_{sc}^* \end{bmatrix} = \begin{bmatrix} I_1^* sin(\omega t) \\ I_1^* sin(\omega t + 120) \\ I_1^* sin(\omega t - 120) \end{bmatrix} \quad (3)$$

PQ distortions are recorded and obtained from customers, from various sources, and the actual source flow using Equation (4), and pulses for the VSI shunt are generated from customer-related PQ distortions using a hysteresis flow controller.

$$\begin{bmatrix} I_{Ca} \\ I_{Cb} \\ I_{Cb} \end{bmatrix} = \begin{bmatrix} I_{sa}^* \\ I_{sa}^* \\ I_{sa}^* \end{bmatrix} - \begin{bmatrix} I_{sa} \\ I_{sb} \\ I_{sc} \end{bmatrix} \quad (4)$$

After connecting the UPFC, the source voltage and current can be found in one phasor together and freed from harmonic distortions using Equation (4). Series parameters and shunt APF can also be obtained from Equations (6) and (7).

$$V_s \prec 0 = Z_s I_s \prec \text{Since } \emptyset_s = 0, \begin{bmatrix} P_s \\ Q_s \end{bmatrix} = \begin{bmatrix} V_s I_s \\ 0 \end{bmatrix} \quad (5)$$

$$V_{sr} \prec (\emptyset_{sr}) = (Z_{sr} + Z_s)I_s \prec 0 = V_R \prec 0 - V_s \prec 0; \begin{bmatrix} P_{sr} \\ Q_{sr} \end{bmatrix} = \begin{bmatrix} V_s I_s cos(\emptyset_{sr}) \\ V_s I_s sin(\emptyset_{sr}) \end{bmatrix} \quad (6)$$

$$I_{sh} \prec \emptyset_{sh} = I_L \prec \emptyset_L - I_s \prec 0, \begin{bmatrix} P_{sh} \\ Q_{sh} \end{bmatrix} = \begin{bmatrix} V_L I_s cos(\emptyset_{sr}) \\ V_s I_s sin(\emptyset_{sr}) \end{bmatrix} \quad (7)$$

Therefore, after compensation, the actual load and the reaction force load are calculated as equations (8) and (9).

$$P_L = V_s I_s \big(1 - cos(\emptyset_{sr})\big) + V_L I_s (cos(\emptyset_{sr}) - cos\emptyset_{sh}) - V_s I_s cos\emptyset_{sh} \quad (8)$$

$$Q_L = V_L I_s (sin(\emptyset_{sr}) - sin\emptyset_{sh}) + V_L I_L sin\emptyset_{sh} - V_s I_s sin(\emptyset_{sr}) \quad (9)$$

Next, it is necessary to apply the neural-fuzzy network or ANFIS to this microgrid system, in order to be able to predict the optimal power distribution as well as the existing hazards for the model to establish a phasor measurement unit based on parallel state estimation. ANFIS is a combined artificial intelligence technique that has the ability to formulate a target map and data using a neural network adaptive control structure and fuzzy logic. In general, the ANFIS architecture has two inputs $(x, y)$ and two rules in the form $(r_1, r_2)$ and the five feedback layers include adaptive or square and non-adaptive or circular. The rules of combined education combine the descending gradient method and estimating the least squares. The adaptive grid under the LSE algorithm, or least squares estimation, assumes that it has an output. In hybrid training algorithms, such as post-diffusion training algorithm, it is possible to optimize the hidden layers and output layer parameters by applying the LSE technique. The hybrid training algorithm not only reduces the dimensions of the search space in the gradient method, but also helps to reduce the time in convergence.

## SIMULATION RESULTS AND ANALYSIS

First, the general settings for the simulation must be provided. The microgrid needs to be connected to the distribution network to the settings that are parametrically entered in Figure 5 to Figure 10. This data is embedded in an Excel file and entered into existing codes and Simulink.

| General Data | |
|---|---|
| Slack | 149 |
| Vnom (kV) | 4.16 |
| InternationalSystem | 0 |
| DeltaLF | 0 |
| V_slack_ph_A | 1.01 |
| V_slack_ph_B | 1.01 |
| V_slack_ph_C | 1.01 |
| Ang_slack_ph_A | 0 |
| Ang_slack_ph_B | -120 |
| Ang_slack_ph_C | 120 |

Figure 5: General Data and Initial Configuration of the Microgrid Connected to the Distribution Network

| Node A | Node B | Length (ft.) | Config. |
|---|---|---|---|
| 149 | 1 | 400 | 1 |
| 1 | 2 | 175 | 10 |
| 1 | 3 | 250 | 11 |
| 3 | 4 | 200 | 11 |
| 3 | 5 | 325 | 11 |
| 5 | 6 | 250 | 11 |
| 1 | 7 | 300 | 1 |
| 7 | 8 | 200 | 1 |
| 8 | 9 | 225 | 9 |
| 14 | 10 | 250 | 9 |
| 14 | 11 | 250 | 9 |
| 8 | 12 | 225 | 10 |
| 8 | 13 | 300 | 1 |
| 9 | 14 | 425 | 9 |
| 34 | 15 | 100 | 11 |
| 15 | 16 | 375 | 11 |
| 15 | 17 | 350 | 11 |
| 13 | 18 | 825 | 2 |
| 18 | 19 | 250 | 9 |
| 19 | 20 | 325 | 9 |
| 18 | 21 | 300 | 2 |

Figure 6: Settings of Nodes Connected to the Microgrid Connected to the Distribution Network and their Configuration Type

| Conf | Lin=1, Trafo=0 | R11 | R12 | R13 | R22 | R23 | R33 | X11 | X12 | X13 | X22 | X23 | X33 | B11 | B12 | B13 | B22 | B23 | B33 |
|---|---|---|---|---|---|---|---|---|---|---|---|---|---|---|---|---|---|---|---|
| 1 | 1 | 0.4576 | 0.1560 | 0.1535 | 0.4666 | 0.1580 | 0.4615 | 1.0780 | 0.5017 | 0.3849 | 1.0482 | 0.4236 | 1.0651 | 5.6765 | -1.8319 | -0.6982 | 5.9809 | -1.1645 | 5.3971 |
| 2 | 1 | 0.4666 | 0.1580 | 0.1560 | 0.4615 | 0.1535 | 0.4576 | 1.0482 | 0.4236 | 0.5017 | 1.0651 | 0.3849 | 1.0780 | 5.9809 | -1.8319 | -1.1645 | 5.3971 | -0.6982 | 5.6765 |
| 3 | 1 | 0.4615 | 0.1535 | 0.1580 | 0.4576 | 0.1560 | 0.4666 | 1.0651 | 0.3849 | 0.4236 | 1.0780 | 0.5017 | 1.0482 | 5.3971 | -0.6982 | -1.1645 | 5.6765 | -1.8319 | 5.9809 |
| 4 | 1 | 0.4615 | 0.1580 | 0.1535 | 0.4666 | 0.1560 | 0.4576 | 1.0651 | 0.4236 | 0.3849 | 1.0482 | 0.5017 | 1.0780 | 5.3971 | -1.1645 | -0.6982 | 5.9809 | -1.8319 | 5.6765 |
| 5 | 1 | 0.4666 | 0.1560 | 0.1580 | 0.4576 | 0.1535 | 0.4615 | 1.0482 | 0.5017 | 0.4236 | 1.0780 | 0.3849 | 1.0651 | 5.9809 | -1.1645 | -1.8319 | 5.6765 | -0.6982 | 5.3971 |
| 6 | 1 | 0.4576 | 0.1535 | 0.1560 | 0.4615 | 0.1580 | 0.4666 | 1.0780 | 0.3849 | 0.5017 | 1.0651 | 0.4236 | 1.0482 | 5.6765 | -0.6982 | -1.8319 | 5.3971 | -1.1645 | 5.9809 |
| 7 | 1 | 0.4576 | 0.0000 | 0.1535 | 0.0000 | 0.0000 | 0.4615 | 1.0780 | 0.0000 | 0.3849 | 0.0000 | 0.0000 | 1.0651 | 5.1154 | 0.0000 | -1.0549 | 0.0000 | 0.0000 | 5.1704 |
| 8 | 1 | 0.4576 | 0.1535 | 0.0000 | 0.4615 | 0.0000 | 0.0000 | 1.0780 | 0.3849 | 0.0000 | 1.0651 | 0.0000 | 0.0000 | 5.1154 | -1.0549 | 0.0000 | 5.1704 | 0.0000 | 0.0000 |
| 9 | 1 | 1.3292 | 0.0000 | 0.0000 | 0.0000 | 0.0000 | 0.0000 | 1.3475 | 0.0000 | 0.0000 | 0.0000 | 0.0000 | 0.0000 | 4.5193 | 0.0000 | 0.0000 | 0.0000 | 0.0000 | 0.0000 |
| 10 | 1 | 0.0000 | 0.0000 | 0.0000 | 1.3292 | 0.0000 | 0.0000 | 0.0000 | 0.0000 | 0.0000 | 1.3475 | 0.0000 | 0.0000 | 0.0000 | 0.0000 | 0.0000 | 4.5193 | 0.0000 | 0.0000 |
| 11 | 1 | 0.0000 | 0.0000 | 0.0000 | 0.0000 | 0.0000 | 1.3292 | 0.0000 | 0.0000 | 0.0000 | 0.0000 | 0.0000 | 1.3475 | 0.0000 | 0.0000 | 0.0000 | 0.0000 | 0.0000 | 4.5193 |
| 12 | 1 | 1.5209 | 0.5198 | 0.4924 | 1.5329 | 0.5198 | 1.5209 | 0.7521 | 0.2775 | 0.2157 | 0.7162 | 0.2775 | 0.7521 | 67.2242 | 0.0000 | 0.0000 | 67.2242 | 0.0000 | 67.2242 |
| 13 | 0 | 0.1000 | 0.0000 | 0.0000 | 0.0000 | 0.0000 | 0.0000 | 0.0000 | 0.0000 | 0.0000 | 0.0000 | 0.0000 | 0.0000 | 0.0000 | 0.0000 | 0.0000 | 0.0000 | 0.0000 | 0.0000 |

Figure 7: One-to-one Configuration Settings of Nodes in the Microgrid Connected to the Distribution Network for Load Distribution

| Node | Y=1, D=0 | Alfa (PQ=0, I=1, Z=2) | Ph-1 (kW) | Ph-1 (kVAr) | Ph-2 (kW) | Ph-2 (kWAr) | Ph-3 (KW) | Ph-3 (kVAr) |
|---|---|---|---|---|---|---|---|---|
| 1 | 1 | 0 | 40 | 20 | 0 | 0 | 0 | 0 |
| 2 | 1 | 0 | 0 | 0 | 20 | 10 | 0 | 0 |
| 4 | 1 | 0 | 0 | 0 | 0 | 0 | 40 | 20 |
| 5 | 1 | 1 | 0 | 0 | 0 | 0 | 20 | 10 |
| 6 | 1 | 2 | 0 | 0 | 0 | 0 | 40 | 20 |
| 7 | 1 | 0 | 20 | 10 | 0 | 0 | 0 | 0 |
| 9 | 1 | 0 | 40 | 20 | 0 | 0 | 0 | 0 |
| 10 | 1 | 1 | 20 | 10 | 0 | 0 | 0 | 0 |
| 11 | 1 | 2 | 40 | 20 | 0 | 0 | 0 | 0 |
| 12 | 1 | 0 | 0 | 0 | 20 | 10 | 0 | 0 |
| 16 | 1 | 0 | 0 | 0 | 0 | 0 | 40 | 20 |
| 17 | 1 | 0 | 0 | 0 | 0 | 0 | 20 | 10 |
| 19 | 1 | 0 | 40 | 20 | 0 | 0 | 0 | 0 |
| 20 | 1 | 1 | 40 | 20 | 0 | 0 | 0 | 0 |
| 22 | 1 | 2 | 0 | 0 | 40 | 20 | 0 | 0 |
| 24 | 1 | 0 | 0 | 0 | 0 | 0 | 40 | 20 |
| 28 | 1 | 1 | 40 | 20 | 0 | 0 | 0 | 0 |
| 29 | 1 | 2 | 40 | 20 | 0 | 0 | 0 | 0 |
| 30 | 1 | 0 | 0 | 0 | 0 | 0 | 40 | 20 |
| 31 | 1 | 0 | 0 | 0 | 0 | 0 | 20 | 10 |

Figure 8: Relationship Values for the Microgrid Connected to the Distribution Network

| Node | Pos X | Pos Y |
|---|---|---|
| 1 | 99.0114 | 323.3191 |
| 2 | 93.1995 | 283.8511 |
| 3 | 99.0114 | 382.8936 |
| 4 | 99.0114 | 401.5106 |
| 5 | 125.7460 | 382.8936 |
| 6 | 156.5490 | 384.3830 |
| 7 | 123.4213 | 318.1064 |
| 8 | 155.3866 | 312.8936 |
| 9 | 145.5064 | 274.9149 |
| 10 | 110.0540 | 271.9362 |
| 11 | 63.5589 | 254.0638 |
| 12 | 140.2757 | 337.4681 |
| 13 | 187.3520 | 306.9362 |
| 14 | 101.3361 | 248.8511 |
| 15 | 206.5312 | 368.7447 |
| 16 | 218.7361 | 397.0426 |
| 17 | 233.8470 | 359.8085 |
| 18 | 148.9936 | 184.0638 |
| 19 | 111.2163 | 193.0000 |

Figure 9: Positions of Micro-Nodes Connected to the Distribution Network on the X and Y Axes

| NODE1 | NODE2 | Closed=1 |
|---|---|---|
| 18 | 135 | 1 |
| 150 | 149 | 1 |
| 13 | 152 | 1 |
| 60 | 160 | 1 |
| 97 | 197 | 1 |

Figure 10: Closed Loop Mode of Nodes in the Microgrid Connected to the Distribution Network

Then, the microgrid block diagram should be modeled for optimal load distribution simultaneously with the establishment and placement of the phasor measurement unit based on parallel state estimation in Simulink environment. The microgrid must be designed. This design is done in Simulink, which is the original and finished model in the form of Figure 11.

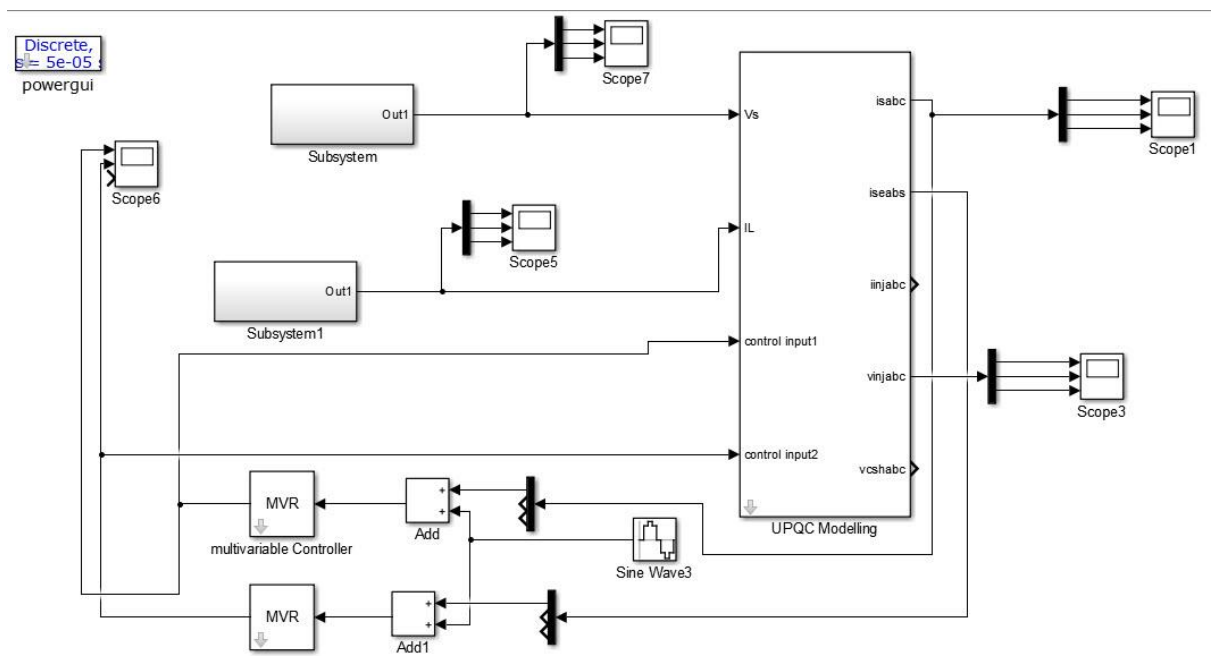

Figure 11: Simulink Microgrid Model for Optimal Load Distribution with UPFC

The main components of this model are created by two variable controllers, fuzzy and neural, which are called MVR. The upper MVR is for the fuzzy part and the lower MVR is for the neural part, which combines to form a neural-fuzzy network and receive commands from the command line when executing existing code. There are subsystems in this Simulink. The subsystem, which is two parts, receives the complete micro-network information once and once as the neural-fuzzy network. The view of this system is as shown in Figure 12.

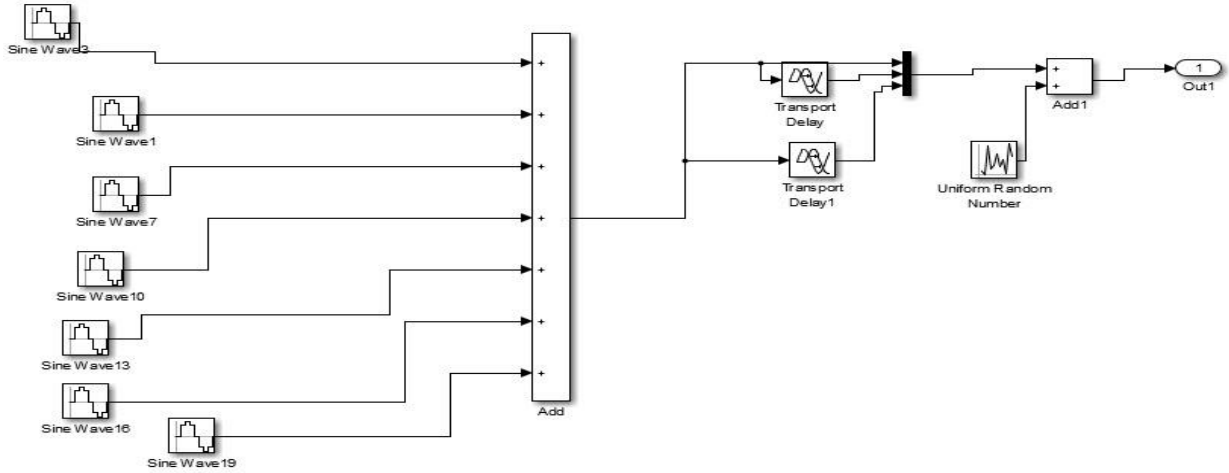

Figure 12: Micro-Network Subsystem to Plot Outputs as Sine Functions

The neural-fuzzy controller is in the form of (13) and is a view of the complete structure of the UPFC in the form of (14) on which the settings are applied based on the initial data of Figures 5 to 10.

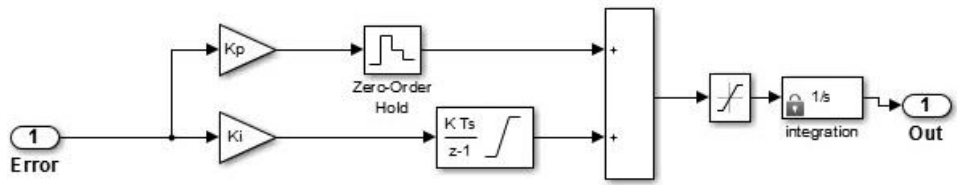

Figure 13: Neuro-fuzzy controller

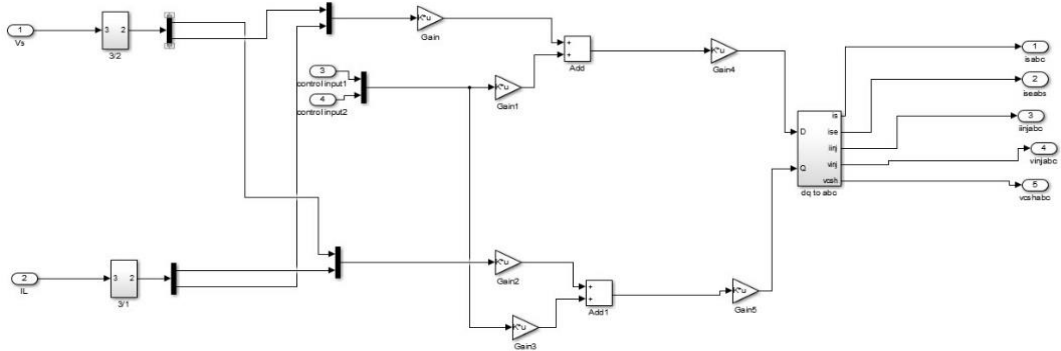

Figure 14: View of the Complete UPFC Structure with Full Settings

When Simulink is running, the source voltage and the voltage load are displayed as outputs, as shown in Figures 15 and 16.

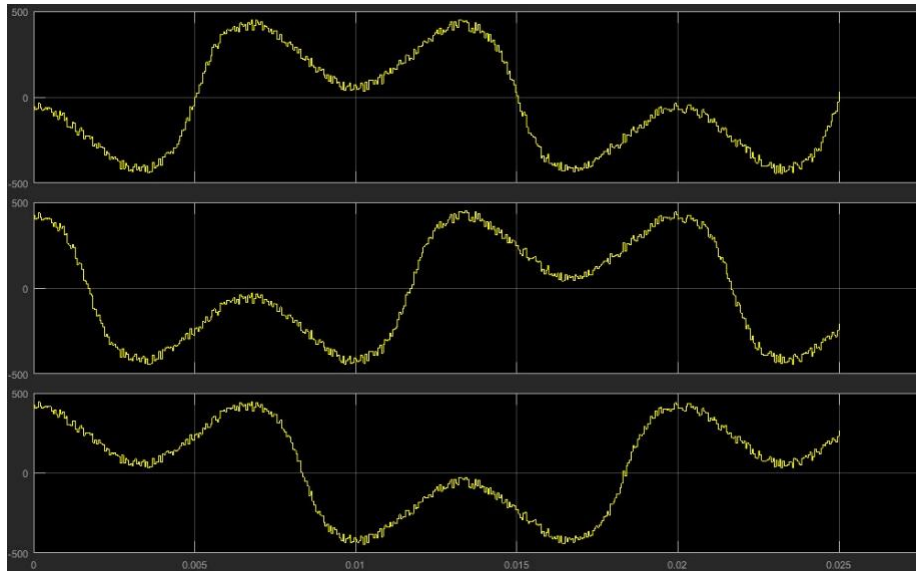

Figure 15: Source Voltage

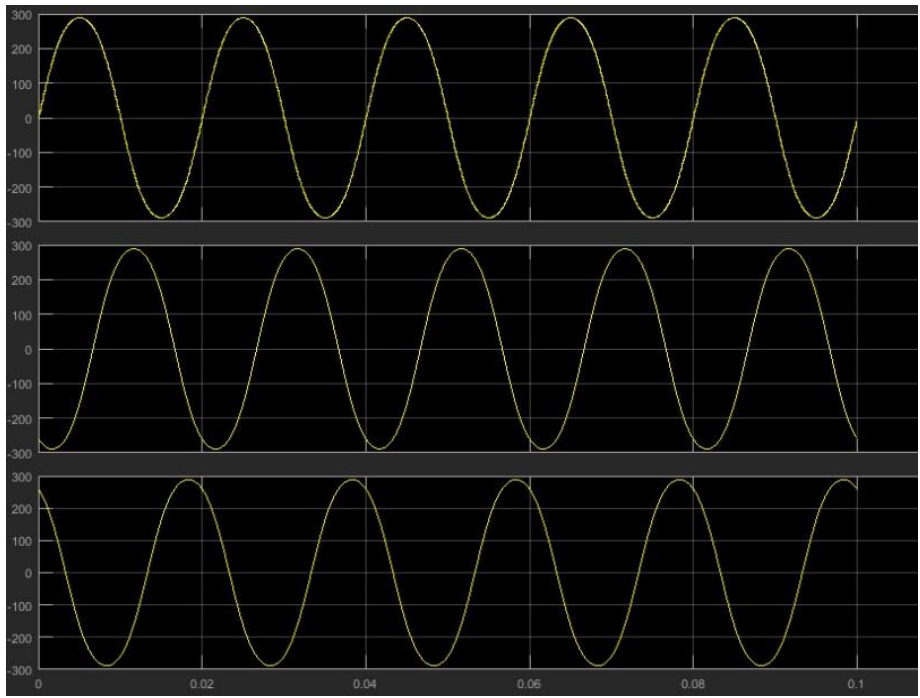

Figure 16: Voltage Load

In general, the data entered for the connections of nodes in the microgrid connected to the distribution network can be shown in Figure 17.

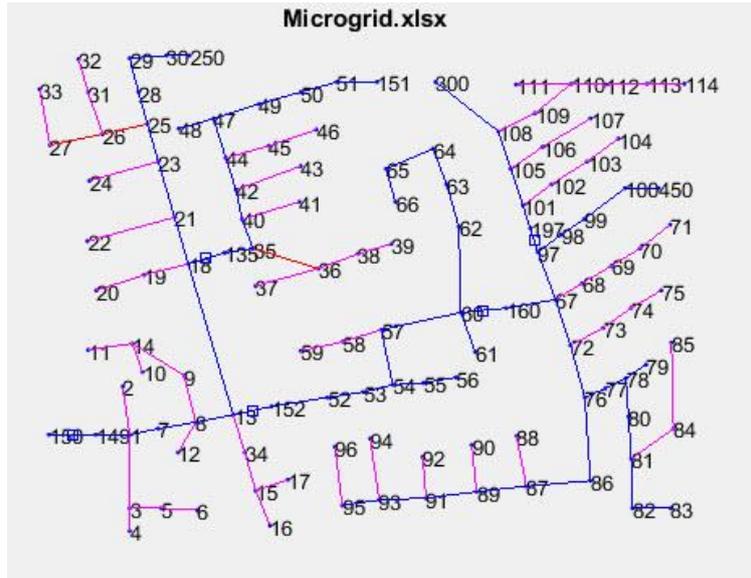

Figure 17: Node Connections in the Microgrid

Also, the errors in the microgrid in terms of percentage during the optimal transfer and distribution of loads in the microgrid level for the establishment and placement of the phasor measurement unit are based on the parallel state estimation as shown in Figure 18.

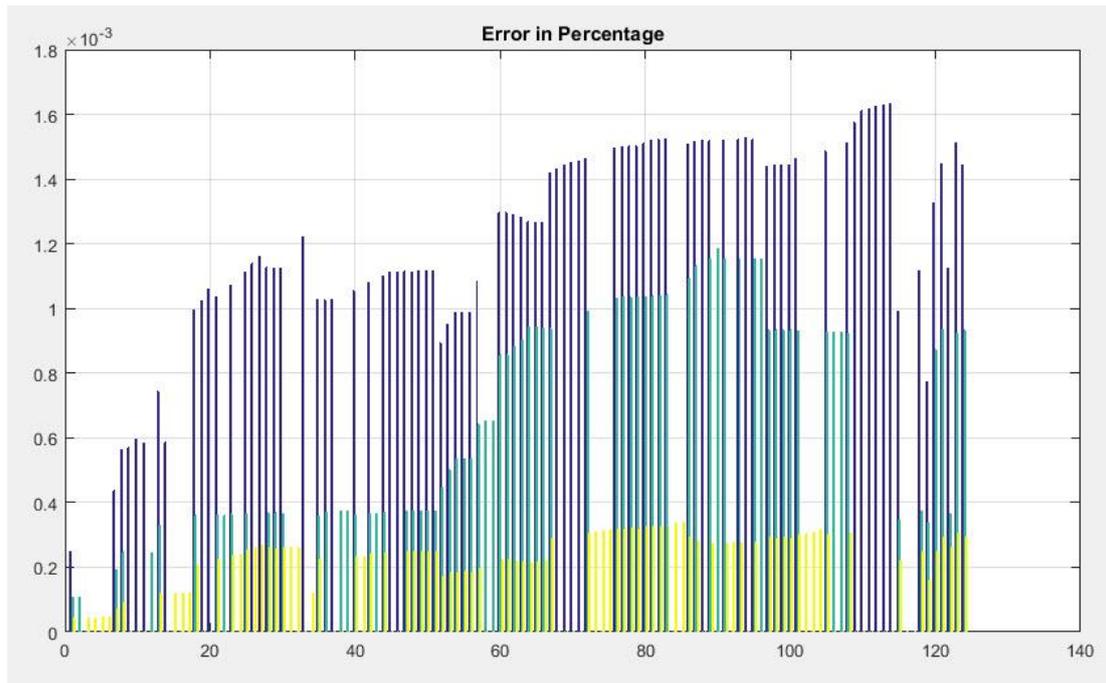

Figure 18: Errors in the Microgrid in Terms of Percentage during Optimal Transmission and Distribution Times

You can also see system analyzes from left to right, including voltage in percent (U), Drop or MGU, power drop rate in U and MGU in percentage, and THD in percentage (U). The output of the analysis is displayed on the MATLAB command line, which is shown in Figure 19.

```
N1    0.9971 < -0.6474    1.0074 < -120.3275    1.0009 <  119.6171
N2    .                   1.0072 < -120.3319    .
N3    .                   .                     0.9993 <  119.5851
N4    .                   .                     0.9988 <  119.5748
N5    .                   .                     0.9980 <  119.5601
N6    .                   .                     0.9974 <  119.5473
N7    0.9876 < -1.1228    1.0056 < -120.5896    0.9951 <  119.3643
N8    0.9814 < -1.4381    1.0043 < -120.7637    0.9912 <  119.1905
N9    0.9799 < -1.4675    .                     .
N10   0.9779 < -1.5070    .                     .
N11   0.9776 < -1.5133    .                     .
N12   .                   1.0040 < -120.7694    .
N13   0.9731 < -1.8756    1.0020 < -121.0084    0.9854 <  118.9030
N14   0.9782 < -1.5005    .                     .
N15   .                   .                     0.9840 <  118.8754
N16   .                   .                     0.9830 <  118.8555
```

Figure 19: Micro-Network System Analysis after Optimal Load Transfer and Distribution

## CONCLUSION

Microgrids connected to distribution networks have many benefits, including improved power quality and reliability, reduced losses, economic benefits, and reduced environmental pollution. Optimal load distribution in distribution networks is an important issue. This necessity stems from the fact that the lack of optimal load distribution can lead to economic and environmental problems on a large scale. Voltage fluctuations are one of the most important problems that can lead to equipment failure. The presence of internal and external disturbances in the microgrids connected to the distribution networks can lead to an increase in frequency in the form of instantaneous events, which exacerbates severe load changes and errors. In microgrids connected to distribution networks, UPFC needs to be used for optimal load distribution, but there are still weaknesses in these systems. Therefore, creating a short-term decision model is essential. It is essential to provide an optimal structure that is capable of training and error detection for the establishment and placement of the phasor measurement unit based on parallel state estimation and further prevention of any internal and external disturbances. Providing control solutions for this is an issue that has been discussed by scientific conferences for many years. Due to the uncertainty in microgrids connected to distribution networks, fuzzy structures can be used, but since there is no possibility of improvement in fuzzy systems, a combined method with it, can be an interesting problem. Therefore, this research uses neural networks along with fuzzy logic, which is called neurophasis or ANFIS. The neural network can automatically create a learning structure during the periodicity in the structure of fuzzy membership functions and generate the opacification operator, which provides an interesting result. After the neural network learning structure, the input signals in the neural network continuously memorize membership functions and fuzzy scientists. Simulations have been performed in MATLAB and Simulink environments, which show that the optimal load distribution can be optimized by considering the frequency and voltage distribution and error detection after the establishment and placement of the phasor measurement unit based on parallel mode estimation. Lead in microgrids connected to distribution networks as a short-term planning model.

**Conflict of Interest Statement**

No conflict of interest has been declared by the authors.